\documentclass{PoS}
\usepackage{epsfig}
\PoS{PoS(LAT2005)114}

\title{The locality of the fourth root of staggered fermion determinant in the interacting case. }

\ShortTitle{The fourth root approximation: interacting theory. }

\author{C. Bernard \\  
      Department of Physics, Washington University, St.~Louis, MO 63130, USA\\
      E-mail: \email{cb@lump.wustl.edu}}

\author{C. DeTar and \speaker{F. Maresca}\\
   Department of Physics, University of Utah, Salt Lake City, UT 84112, USA\\
   E-mail: \email{detar@physics.utah.edu}, \email{maresca@physics.utah.edu}}

\author{Steven Gottlieb and L. Levkova\\
Department of Physics, Indiana University, Bloomington, IN 47405, USA \\
E-mail: \email{sg@physics.indiana.edu}, \email{llevkova@indiana.edu}}

\author{U.M. Heller\\
 American Physical Society, One Research Road, Box 9000, Ridge, NY 11961, USA\\
 E-mail: \email{heller@aps.org}}

\author{J.E. Hetrick\\
  Physics Department, University of the Pacific, Stockton, CA 95211, USA\\
 E-mail: \email{jhetrick@pacific.edu}}

\author{R. Sugar\\
Department of Physics, University of California, Santa Barbara, CA 93106, USA\\
E-mail: \email{sugar@vulcan2.physics.ucsb.edu}}

\author{D. Renner and D. Toussaint\\ 
     Department of Physics, University of Arizona, Tucson, AZ 85721, USA\\
E-mail: \email{doug@physics.arizona.edu}, \email{dru@physics.arizona.edu}}

\abstract{The fourth root approximation in LQCD simulations with
dynamical staggered fermions requires justification. We test its validity
numerically in the interacting theory in a renormalization group
framework.}

\FullConference{XXIIIrd International Symposium on Lattice Field Theory\\
             25-30 July 2005\\
             Trinity College Dublin, Ireland}

\begin{document}

\section{Introduction}
Numerical simulations of dynamical fermions within the framework of 
the staggered formalism are both computationally 
cost effective and  phenomenologically successful.
Exploiting the advantageous properties of an improved staggered fermion
formulation, various collaborations are performing high-precision lattice QCD 
calculations that are in excellent agreement with experimentally known 
measurements \cite{MILC}. 
However, this success is clouded by the long-standing problematic issue
of the validity of the fourth root approximation.
The staggered fermion describes, in fact, four tastes on the lattice, so
in order to study QCD with $N_f = 2$ or $N_f = 2+1$, a Boltzmann weighting
is used that contains a fractional power of the fermion determinant.
Because of taste-breaking at nonzero lattice spacing, 
taking the fractional power of the staggered determinant
before restoring taste symmetry is conceptually nontrivial.
The issue whether the fourth root prescription gives a lattice theory
in the right universality class to reproduce QCD is yet unresolved and
the phenomenological success of the staggered formulation has heated up the 
discussion. In the past years various numerical investigations 
\cite{Durr} have addressed this concern.
Recently a direct approach was adopted 
towards establishing whether the universality class is the right one: 
if a single-taste local fermion action can be found, whose
determinant is equal to the fourth root of the staggered fermion
determinant up to cutoff effects
\begin{equation}\label{decomp}
\lim_{a \rightarrow 0} (\mbox{ det } D_{st})^{1/4} = \mbox{det }D  \ \mbox{det } H
\end{equation}
(where $D$ is a local Dirac operator and $H$ is local and contains only cutoff 
effects) then the fourth root prescription 
can be consistently fitted into the framework of a local field theory.
In the free theory numerical \cite{Peardon} and analytical
 studies \cite{Adams,Shamir} showed that such a local operator exists.
In particular, in \cite{Shamir} Shamir applied renormalization group 
blocking to the free staggered operator $D_0$ in the spin 
$\otimes$ taste representation. 
After $n$ blocking steps $Q_n$, the fermionic degrees of freedom live 
on the coarse lattice with lattice spacing $a_c = 2^n a$ and the determinant 
of the staggered operator decomposes as 
$\mbox{det }D_{st}(a = 2^{-n} a_c) = \mbox{det }D_n \ \mbox{det }G_n^{-1}$.
More details can be found in \cite{Shamir}, where it was proved 
analytically that the blocked
propagator $D^{-1}_n = \alpha^{-1} + Q_n D_0^{-1}Q_n^\dagger$ factorizes, in 
the limit $n \rightarrow \infty$, as 
$D^{-1}_{\infty} = D_{\mbox{rg}} \otimes I$ and that  
$G^{-1}_n  =  D_0 + \alpha \ Q_n^\dagger Q_n$ is a local operator
that contains only cutoff effects. This completes the proof that in the
continuum limit, \textit{i.e.} $ a (=2^{-n} a_c) \rightarrow 0$,
 the decompositions of Eq.~(\ref{decomp}) holds in the free case.       

\section{Renormalization group transformation in the interacting case}
In this section we briefly discuss the renormalization group (RG)
program adopted to test the interacting case.  
The new complication is that a mapping from the one-component staggered
fermion basis to the spin $\otimes$ taste basis is not
unique. We define the interacting theory in the 
one-component formalism.  The first fermion RG blocking 
transformation is used to perform the transition to the 
spin $\otimes$ taste representation:
\begin{equation}
(Q_{(1)} \psi)^{a \alpha}(2x) = \frac{1}{2}\sum_{r_\mu=0,1}( \gamma_1^{r_1} \gamma_2^{r_2}  \gamma_3^{r_3}  \gamma_4^{r_4} )^{(a \alpha)} \ W_r(2x,2x+r) \psi(2x+r) ,
\end{equation}
where $a$ and $ \alpha$ are Dirac spin and taste indices respectively. 
We require the parallel transporters $W_r(2x,2x+r)$ to be a sum over all 
the shortest paths~\footnote{These are built using APE smeared links of the 
original lattice to be consistent with the RG program adopted 
for the subsequent RG steps.}  between the origin  $2x$ and the other sixteen sites $2x + r$ of the hypercube.
The sum over different paths is thought 
to reduce the breaking of  hypercubic symmetry; however, the choice to 
parallel transport all the points within a given hypercube to the 
 corner $r = 0$ unavoidably introduces some asymmetries.        
The subsequent RG steps are a covariant generalization of the
arithmetic mean on a $2^4$ hypercube used in \cite{Shamir} :
\begin{equation}\label{rg2}
(Q_{(n)} \psi)(2x) = 2^{-4} \sum_{r_\mu=0,1} W_r(2x,2x+r) \psi(2x+r) \hspace{4ex} n > 1 .
\end{equation}
The shortest paths that compose the parallel transporters  $W_r(2x,2x+r)$ in
Eq.~(\ref{rg2}) are constructed from links of the blocked lattice (with lattice 
spacing $2^n \ a$). These blocked links are built following  a program 
suggested in \cite{Degrand}. In detail, the links that live on the fine 
lattice  (with lattice spacing $2^{n-1} \ a$) are twice APE smeared and 
projected back to $SU(3)$. 
The blocked links are then built multiplying two of 
these fine smeared links in line.
Once the blocking kernels are defined we project the nondiagonal piece of the 
blocked propagator $D^{-1}_{n}$ onto the spin $\otimes$ taste Clifford space:
\begin{equation}\label{proj}
(Q_n D_0^{-1} Q^\dagger_n)(x,y=0) = \sum_{S,T} \Gamma_S \otimes \Gamma_T^\dagger \ M_{ST}(x,y=0) .
\end{equation}
Here $x$ and $y=0$ \footnote{The fermion source is defined at the origin.} live on the coarsest lattice and $Q_n$ 
can be regarded as a 'big'  blocking step 
$Q_n=Q_{(n)} Q_{(n-1)} \cdots  Q_{(1)}$ that transforms the original lattice 
with lattice spacing $a$ 
into the blocked lattice with lattice spacing $a_c = 2^n a$.    
The coefficients $M_{ST}(x,y=0)$ of the projection in Eq.~(\ref{proj})
are evaluated numerically and averaged over two different lattice 
ensembles.  We fixed Lorentz gauge before doing any RG blocking.
Parameters of the simulations are shown in Table \ref{table1}.
Applying a different number of RG steps, $n$, on each of these two 
ensembles, the resulting blocked lattices have the same coarse-lattice spacing 
of  $a_c = 0.72$ fm. 
This allowed us to see how the coefficients $M_{ST}$ scale with one
additional RG blocking step.

\begin{table}[h]
\begin{center}
\begin{tabular}{|l|l|l|}
\hline
unblocked lattice & $16^3 \times 48$  &  $40^3 \times 96$ \\
\hline
$a$ & 0.18 fm  & 0.09 fm \\
\hline
$m_s \ a $ &  0.125  & 0.05 \\
\hline
number of RGT's &  2 &  3 \\
\hline
blocked lattice & $4^3 \times 12$  &  $5^3 \times 12$ \\
\hline
$a_c$ & 0.72 fm  & 0.72 fm \\
\hline
number of cfgs & 148 & 56 \\
\hline
\end{tabular}
\caption{Simulation parameters}
\label{table1}
\end{center}
\end{table}

\begin{figure}[!h]
\begin{center}
\epsfig{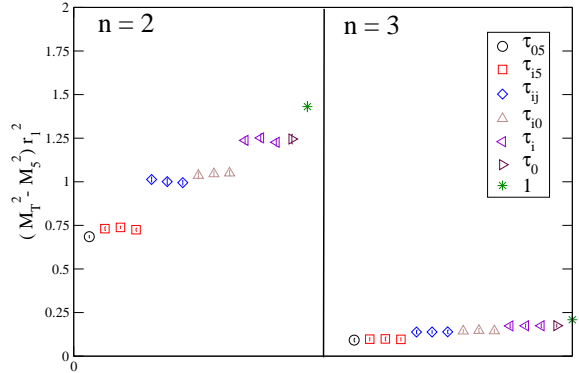}
\caption{Pseudoscalar taste splittings after 2 and 3 RG blockings in units
of $r_1$, the distance at which the static quark potential satisfies
$r^2dV/dr = 1$.}
\label{figmass}
\end{center}
\end{figure}

\vspace{-2ex}
As a consistency check on the blocking procedure described above, we evaluated
the mass of the $(s \overline{s})$ meson.
%\footnote{We fixed lattice Lorentz 
%gauge before doing any RG blocking.}. 
In fact, the spectrum
on the blocked lattice is expected to be identical to the 
spectrum in the original theory. The meson mass can be 
evaluated directly from the coefficients $M_{ST}(x,y)$ of Eq.~(\ref{proj}):
\vspace{-1ex}
\begin{equation}
C^{T}_{s\overline{s}} (t) = \sum_{S,T^\prime,\vec{x}} < \mbox{Tr} (\tau_5 \Gamma_{T} \Gamma_{T^\prime} \Gamma_{T} \tau_5 \Gamma_{T^\prime}^{\dagger}) \ |M_{ST^\prime}(t,\vec{x})|^2>,
\end{equation}
where the $\Gamma_{S}$'s and the $\Gamma_{T,T^\prime}$'s are the 16 Dirac 
matrices in the normal and adjoint ($\tau_{\mu} = \gamma_\mu^{\star}$) 
representation respectively.    
In Figure \ref{figmass} we present the taste splittings for the meson masses
evaluated on the two blocked lattices.
At this level of statistics, the decrease in splitting from $n = 2$ to
$n = 3$ is consistent with the expected ${\cal O}(a^2 \alpha)$ or
${\cal O}(a^2 \alpha^2)$, but the preliminary RG blocked splittings are
systematically higher than values obtained in direct measurements at
much lighter valence quark masses.  Further study is needed.
 
\vspace{5mm}

\vspace{-4ex}

\section{Preliminary results}

In this preliminary study we have analyzed for various displacements $|x-y|$
the dominant terms in Eq.~(\ref{proj}) that are also present in the free theory:
\begin{equation}\label{proj1}
\sum_\mu  \ (\gamma_\mu \otimes I )  A^{(n)}_\mu(x,y) + ( \gamma_5 \otimes \tau_5 \tau_\mu ) B^{(n)}_\mu (x,y) +  
\sum_{\nu \neq \mu}( \gamma_5 \otimes \tau_5 \tau_\nu ) B^{\prime \ (n)}_\nu(x,y) + ( I \otimes I ) \ C^{(n)}(x,y) +  \cdots \cdots 
\end{equation}

\begin{figure}[!h]
\begin{minipage}{7cm}
\begin{center}
\epsfig{file=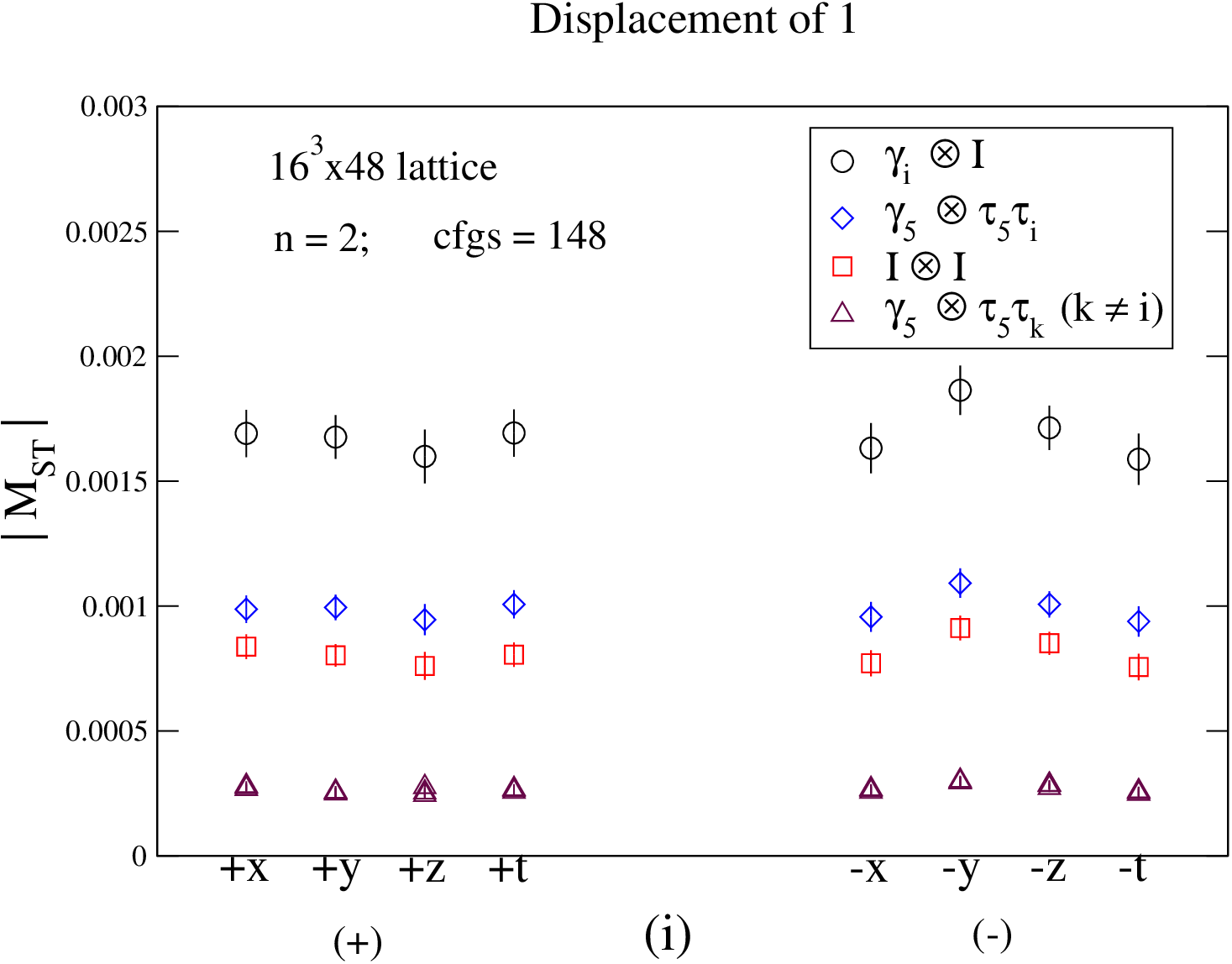, width=1.0\columnwidth}
\caption{$M_{ST}$ after $n=2$ RG steps. }
\label{fig1}
\end{center}
\end{minipage}
\qquad
\begin{minipage}{7cm}
\epsfig{file=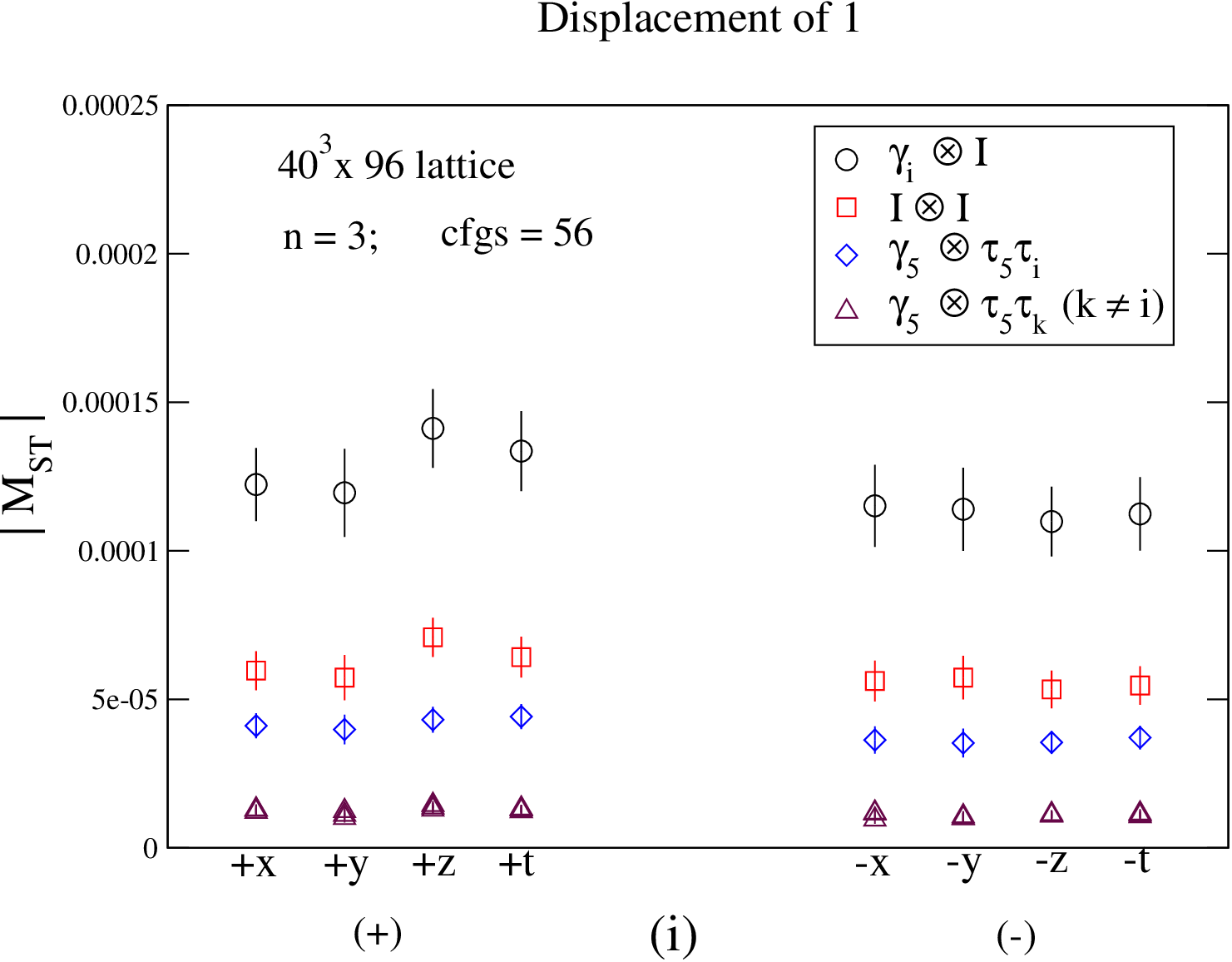, width=1.0\columnwidth}
\caption{$M_{ST}$ after $n=3$ RG steps.}
\label{fig2}
\end{minipage}
\end{figure}

Comparing Figures \ref{fig1} and \ref{fig2}, 
it is remarkable to see that increasing by only one unit the number of 
RG blocking transformations the mass term $( I \otimes I )$ 
( {\color{red} $ \square$} ) becomes larger than the taste 
breaking term  $\gamma_5 \otimes \tau_5 \tau_i $ 
( {\Large{{\color{blue} $ \diamond$}}} ) for any direction $i$ for displacement 
$|x-y|=1$. This is in agreement with what we expect from
the free theory: in the limit $n \rightarrow \infty$ the only terms 
in $M_{ST}$ 
that should survive are diagonal in taste space.  

Other terms in Eq.~(\ref{proj1}) are seen in our simulations. 
Their contribution is $10-1000$ times smaller than the leading kinetic 
term $A^{(n)}_\mu$ in Eq.~(\ref{proj1}) 
and this make it difficult to identify the   
terms that are statistically relevant.  We need to increase the number of 
configurations to have a better understanding of the statistics.
Besides this issue, it is crucial to know how these terms scale with 
$n$ to determine whether the blocked propagator 
becomes diagonal in taste space in the interacting theory as it does in the 
free theory. However an element that can play an important role in this 
scaling is the breaking of the hypercubic symmetry, caused by the introduction
of the parallel transporters, as it has been briefly discussed in the previous 
section. In fact, Figures \ref{fig1} and \ref{fig2} suggest that the blocking procedure
is introducing some hypercubic asymmetries, since the coefficients 
$M_{ST}(x,y)$ are not perfectly degenerate under reflections
 about the principal axes.
In order to investigate further these terms, present only
in the interacting theory, it is necessary either to quantify the 
hypercubic symmetry breaking or to redefine the RG blocking transformations
in order to guarantee hypercubic invariance (one suggestion is presented 
at this conference \cite{Shamir05}).        
 
%\begin{figure}[!h]
%\begin{center}
%\epsfig{file=16disp11.eps, width=.5\textwidth}
%\caption{$M_{ST}(x,y)$'s under rotations and reflections about the principal axes}
%\label{fig3}
%\vspace*{-4mm}
%\end{center}
%\end{figure}

\section{Interacting theory \textit{vs} free theory}

\begin{figure}[!ht]
\begin{minipage}{7cm}
\begin{center}
\epsfig{file=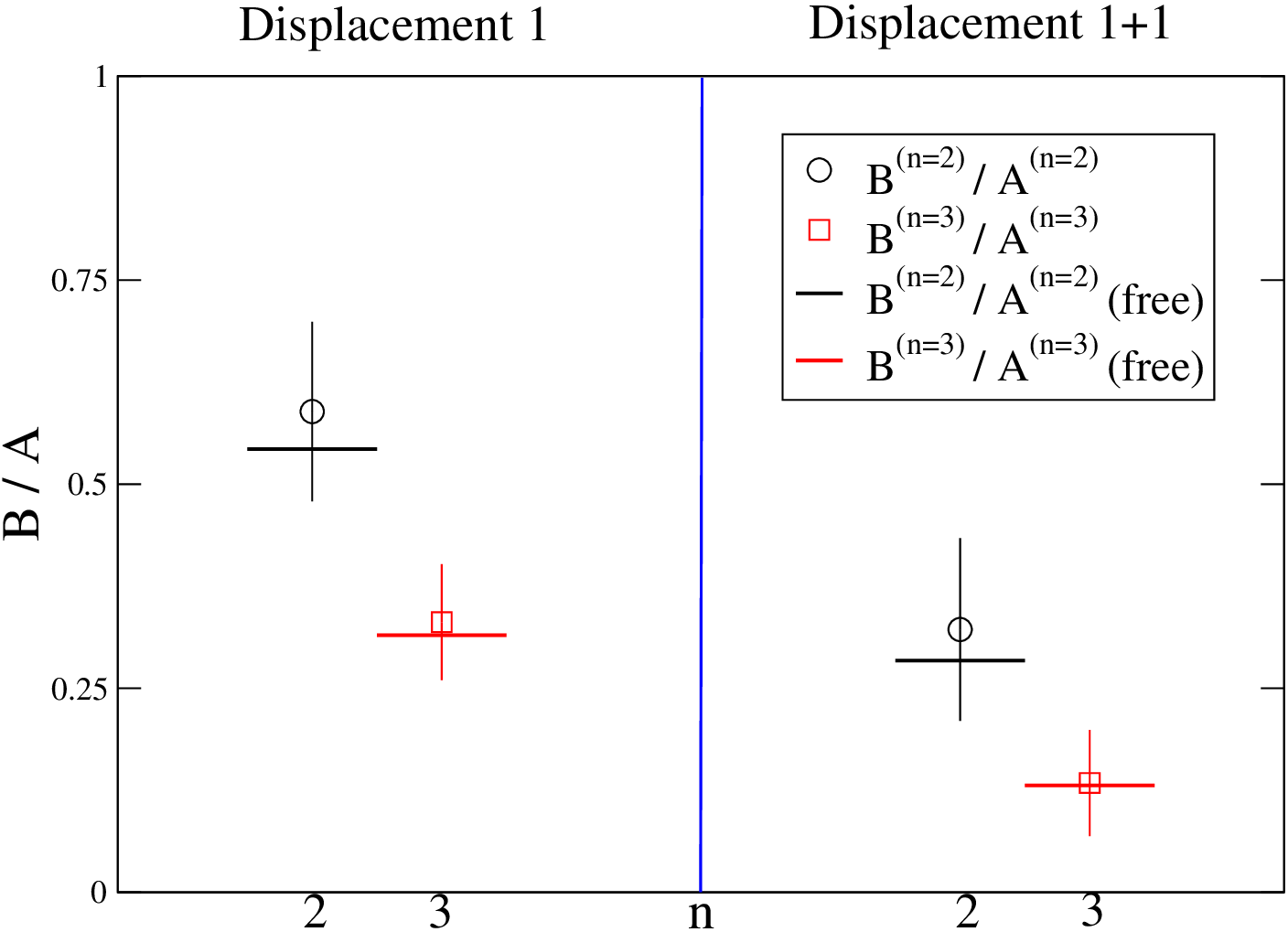, width=1.0\columnwidth}
\caption{Taste breaking term $B^{(n)}$ over the kinetic term. }
\label{fig4}
\end{center}
\end{minipage}
\qquad
\begin{minipage}{7cm}
\epsfig{file=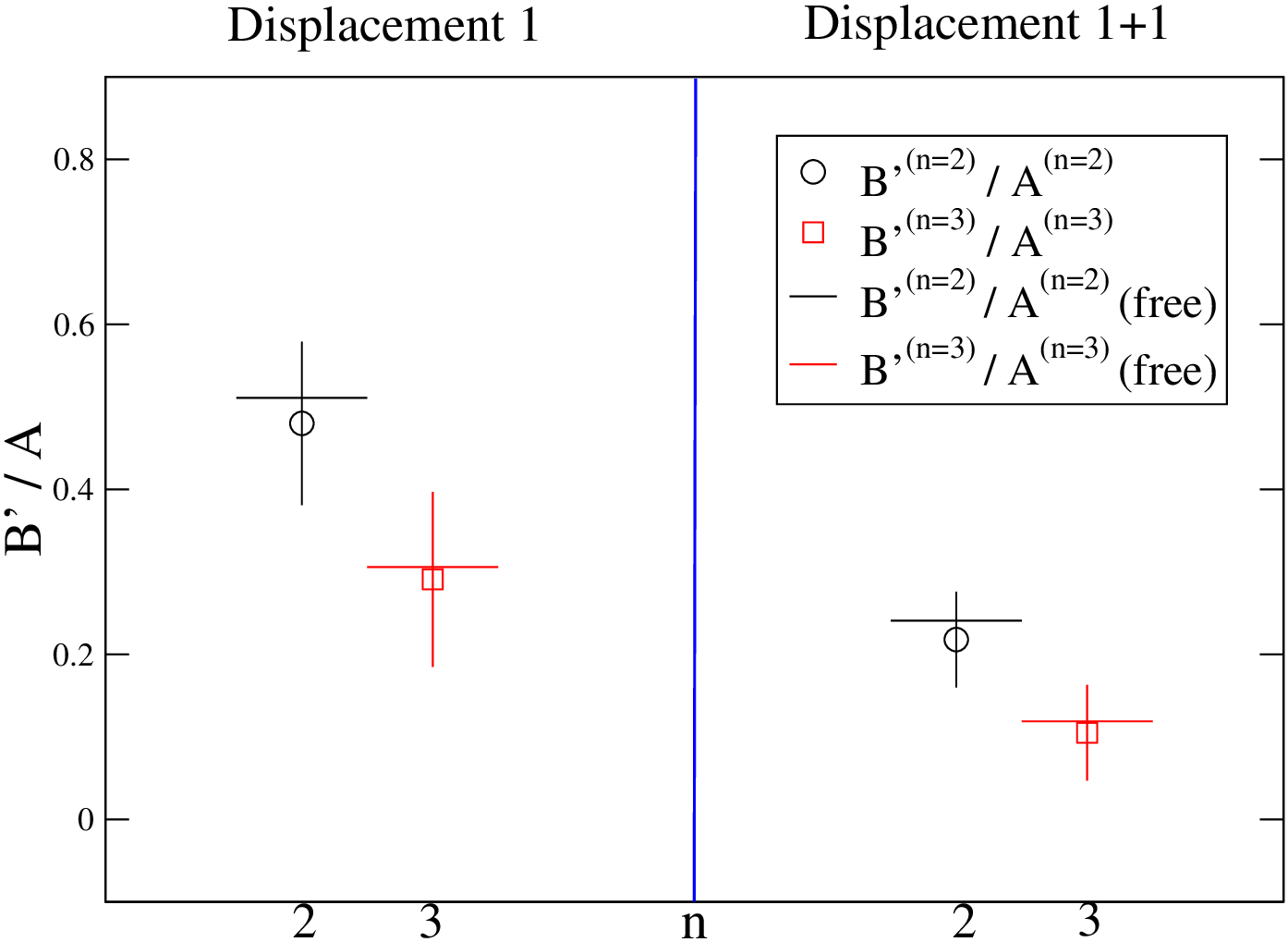, width=1.0\columnwidth}
\caption{Taste breaking term $B^{\prime (n)}$ over the kinetic term.}
\label{fig5}
\end{minipage}
\end{figure}

In \cite{Shamir} the scaling properties of the taste breaking terms 
$B^{(n)}_\mu(p)$  were evaluated analytically.
It was shown in momentum space that they  scale like $2^{-n}$, 
uniformly in $p$, so the blocked propagator becomes diagonal
in taste space, when $n \rightarrow \infty$. 
Postponing the issue of the breaking of hypercubic symmetry to a later
study, we consider here a linear combination of the amplitudes $|M_{ST}(x,y)|$
 under reflections and rotations about the principal axes.  
In Figures \ref{fig4} and \ref{fig5} we show how the ratios between the
taste violating terms and the leading kinetic term  $A^{(n)}$ 
scale with $n$. 

\vspace{1mm}
\vspace{-2ex}

\begin{figure}[!h]
\begin{minipage}{7cm}
\begin{center}
\epsfig{file=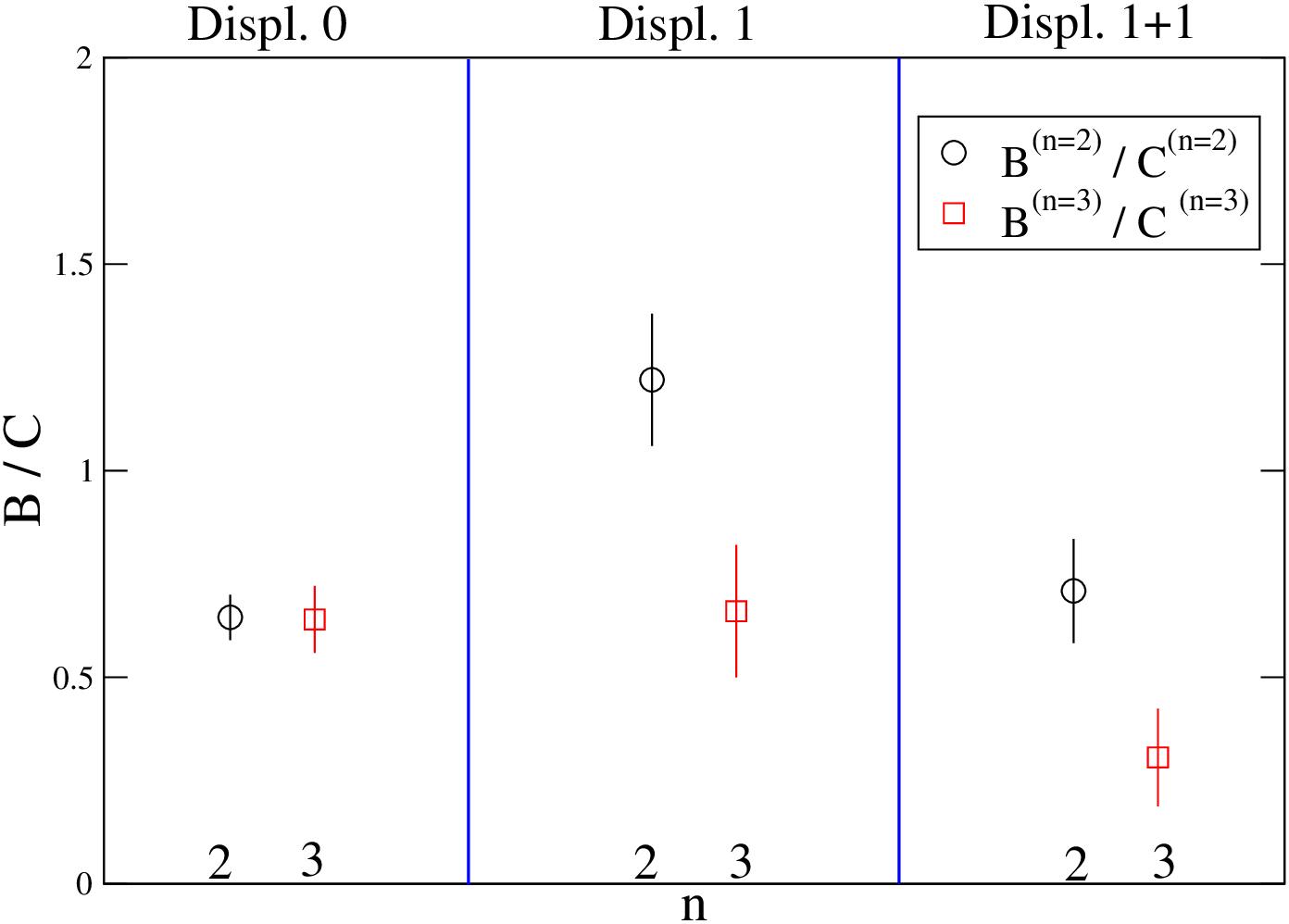, width=1.0\columnwidth}
\caption{Taste breaking term divided by the mass term: interacting theory }
\label{fig6}
\end{center}
\end{minipage}
\qquad
\begin{minipage}{7cm}
\epsfig{file=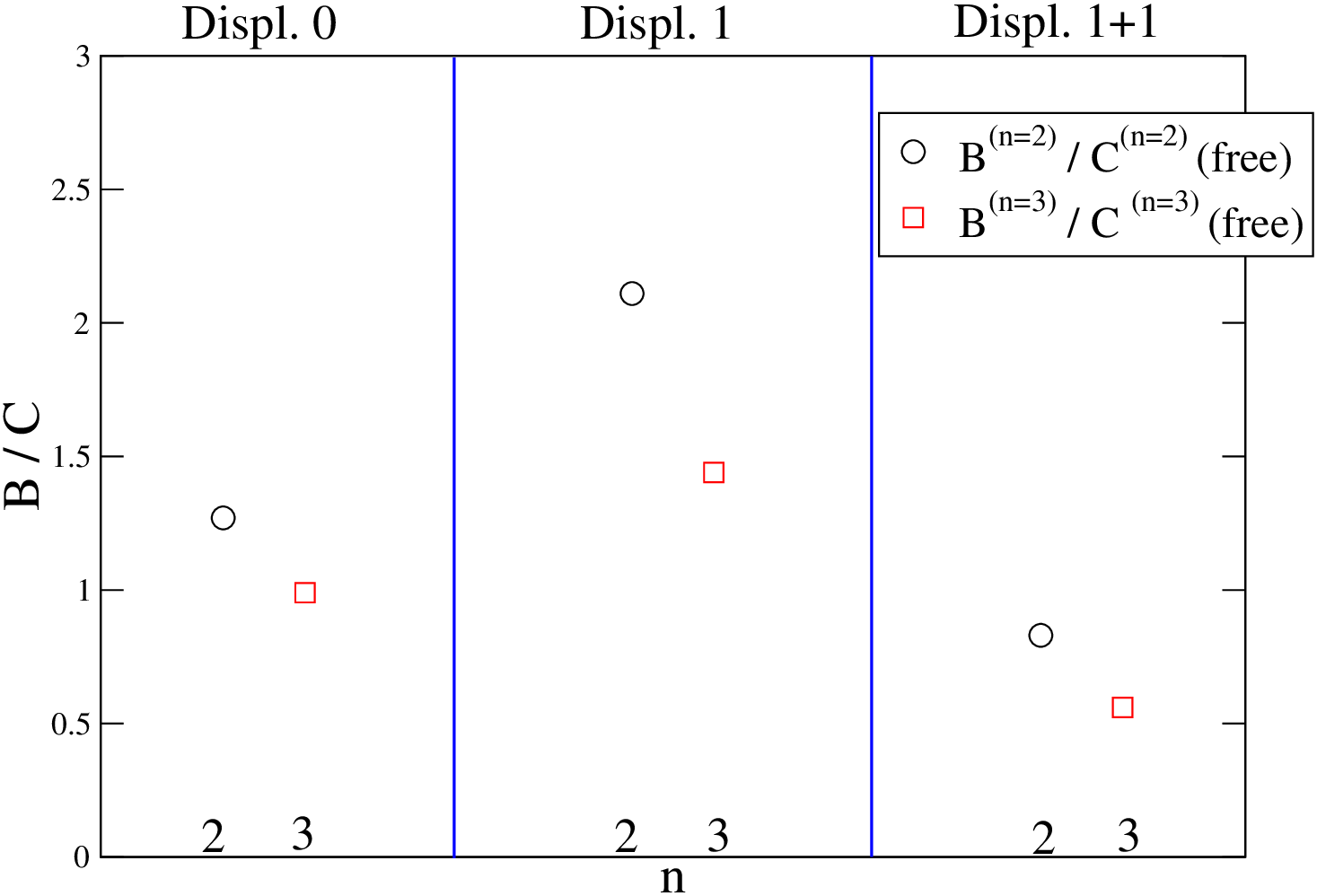, width=1.0\columnwidth}
\caption{Taste breaking term divided by the mass term: free theory.}
\label{fig7}
\end{minipage}
\end{figure}

Our data show that they diminish when $n$ is increased  and 
 that their magnitudes are also remarkably in agreement with the free theory 
represented by solid lines.
In Figures \ref{fig6} and \ref{fig7} we show the scaling properties of the 
taste breaking term divided by the mass term. 
The interacting theory agrees with the 
free theory for displacements $|x-y|=1,2$, however a discrepancy is seen 
at zero displacement where the interacting theory does not scale as expected.
The zero displacement is not interesting when discussing locality 
 thus this behavior, even if not well understood, does not spoil the good 
scaling properties we are seeing in the interacting theory.

\vspace{2mm}
\vspace{-2ex}

The conclusion that we can draw from this preliminary study is that 
the blocked propagator in the interacting case is dominated by the same
 terms present in the free theory and that these terms scale as expected to 
make the decomposition of Eq.~(\ref{decomp}) possible. The statistical relevance  and the scaling properties of other terms seen in the interacting theory 
is still under investigation.

We thank Yigal Shamir and Maarten Goltermann for helpful discussions
during the conference.  Computing support from the Utah Center for
High Performance Computing is gratefully acknowledged.  This work was
supported by the US NSF and DOE.

\end{document}